# Classifying charge carrier interaction in highly-compressed elements and silane

Evgeny F. Talantsev[1,2]

[1]M.N. Mikheev Institute of Metal Physics, Ural Branch, Russian Academy of Sciences, 18, S. Kovalevskoy St., Ekaterinburg, 620108, Russia

[2]NANOTECH Centre, Ural Federal University, 19 Mira St., Ekaterinburg, 620002, Russia

**Abstract**

Since pivotal experimental discovery of the near-room-temperature superconductivity (NRTS) in highly-compressed sulphur hydride by Drozdov *et al* (2015 *Nature* **525** 73-76), more than a dozen of binary and of ternary hydrogen-rich phases exhibited superconducting transition above 100 K have been discovered to date. There is a widely accepted theoretical point of view that primary mechanism governing the emergence of superconductivity in hydrogen-rich phases is the electron-phonon pairing. However, recent analysis of experimental temperature dependent resistance, $R(T)$, in $H_3S$, $LaH_x$, $PrH_9$ and $BaH_{12}$ (Talantsev 2021 *arXiv*: 2104.14145) showed that these compounds exhibit the dominance of non-electron-phonon charge carrier interaction and, thus, it is unlikely that the electron-phonon pairing is the primary mechanism for the emergence of superconductivity in these materials. Here we use the same approach to reveal charge carrier interaction in highly-compressed lithium, black phosphorous, sulfur, and silane. We found that all these superconductors exhibit the dominance of non-electron-phonon charge carrier interaction. This explains the failure of high-$T_c$ values predicted for these materials by the first-principles calculations which utilized the electron-phonon pairing as the mechanism for the emergence of superconductivity in these materials. Our result implies that alternative pairing mechanisms (i.e., electron-magnon, electron-polaron, electron-electron, etc.) should be tested within first-principles calculations approach as possible mechanisms for the emergence of superconductivity in highly-compressed superconductors.

**Classifying charge carrier interaction in highly-compressed highly-compressed elements and silane**

**I. Introduction**

Th$_4$H$_{15}$ was the first superhydride which was discovered by Satterthwaite and Toepke [1] based on their pivotal idea [1]: "…There has been theoretical speculation [2] that metallic hydrogen might be a high-temperature superconductor, in part because of the very high Debye frequency of the proton lattice. With high concentrations of hydrogen in the metal hydrides one would expect lattice modes of high frequency and if there exists an attractive pairing interaction one might expect to find high-temperature superconductivity in these systems also."

Nearly twenty superconducting superhydride phases have been discovered [3-20] since milestone report by Drozdov *et al* [3] on the observation of the superconducting transition above 200 K in highly-compressed sulfur hydride H$_3$S [3]. Despite widely acknowledge consensus, supported by the first-principles calculations, that primary mechanism governs the near-room-temperature superconductivity (NRTS) in superhydrides is the electron-phonon pairing, there are several dramatic failures of this approach.

For instance, we can highlight the prediction by Feng *et al* [21] who calculated the Debye temperature of $T_\theta = 3{,}500 - 4{,}000\ K$ and $T_c \cong 165\ K$ for highly-compressed hydrogen-rich silane, SiH$_4$, for which the experiment performed by Eremets *et al* [22] showed the onset of superconducting transition $T_c^{onset} = 7 - 17\ K$ at a pressure varied in a wide range of $60\ GPa \leq P \leq 192\ GPa$. Talantsev [23] deduced the Debye temperature of $T_\theta = 353 \pm 3\ K$ by fitting experimental $R(T)$ data for SiH$_4$ (compressed at $P = 192\ GPa$ and exhibited $T_c^{onset} \cong 11\ K$) to the Bloch-Grüneisen (BG) equation [24,25]:

$$R(T) = R_0 + A \cdot \left(\frac{T}{T_\theta}\right)^5 \cdot \int_0^{\frac{T_\theta}{T}} \frac{x^5}{(e^x-1)\cdot(1-e^{-x})} \cdot dx \qquad (1)$$

where, $R_0$ is the residual resistance at $T \rightarrow 0\ K$ due to the scattering of electrons (this type of charge carrier was confirmed in direct experiment for NRTS H₃S [26]) on the static defects of the crystalline lattice, and the second term describes the electron-phonon scattering, where $A$ and $T_\theta$ are free-fitting parameters. Thus, experimentally observed $T_c^{onset} \cong 11\ K$ and deduced $T_\theta = 353 \pm 3\ K$ are well agreed with the weak-coupling scenario, $\frac{T_c^{onset}}{T_\theta} = 0.03$. But, both these values are different from computed ones [21] by more than one order of magnitude.

It should be stressed that this failure of the first-principles calculations for one of the chemically simplest hydrogen-rich superconductors is still at the same unexplained and uncommented status since 2008 [21,22,27,28].

There are several nearly identical failures of the first-principles calculations for low-Z elements, for which we can mention highly-compressed lithium, for which predicted $T_c = 50 - 90\ K$ [29] with exact predicted $T_c = 55\ K$ at $P = 40\ GPa$ [29]. Experiments show a small drop (about 5% from normal state resistance) at $T_c^{onset} \sim 7\ K$ at a pressure of 22 GPa $\leqslant P \leqslant$ 32 GPa [30]. Shimizu *et al* [31] reported also a small drop in resistance $T_c^{onset} \sim 6\ K$ at $P = 40\ GPa$ [31], while Struzhkin *et al* [32] reported the diamagnetic signal at $T_c^{onset} \sim 10\ K$. It should be mentioned, that for one sample, compressed at $P$ = 48 GPa, Shimizu *et al* [31,33] reported $T_c^{onset} \cong 20\ K$, while other samples exhibited $6\ K \leq T_c^{onset} \leq 10\ K$ at a wide pressure range of $23\ GPa \leq P \leq 80\ GPa$. Deemyad *et al* [34], Matsuoka *et al* [35] and Schaeffer *et al* [36] reported $T_c^{onset} \leq 14\ K$ for lithium compressed at a pressure range of $16\ GPa \leq P \leq 60\ GPa$. Even the most advance first-principles calculations [37] did not reproduce *a priori* known experimental $T_c^{onset}(P > 30\ GPa)$ dataset for highly-compressed lithium. Extended review of the problem for other materials can be founded elsewhere [23].

However, it should be stressed that the electron-phonon pairing [38,39] is not the only mechanism which can cause the formation and the condensation of the Cooper pairs. For instance, recently Kim [40] propose a theory of NRTS based on the electron-electron retractive pairing. This theory is in a good accord with recent report by Talantsev [41] who analysed temperature dependence of the resistivity, $R(T)$, in superconducting highly-compressed superhydrides $H_3S$, $LaH_x$, $PrH_9$ and $BaH_{12}$ and reported that all these materials exhibit the dominance of non-electron-phonon charge carrier interaction.

Here we applied the same approach as in Ref. 41 to reveal the dominant charge carrier interaction in highly-compressed lithium, black phosphorous, sulfur, and silane. In the result, we found that all these superconductors also exhibit the dominance of non-electron-phonon charge carrier interaction. And, thus, at least partially, the failure of the first-principles calculations based on a presumption of electron-phonon mediated superconductivity in some of these materials can be explained.

## 2. Model description

Jiang *et al* [42] and later Talantsev [41,43] proposed to reveal the type of the charge carrier interaction in metallic substances using a generalized version of the Bloch-Grüneisen equation [41-43]:

$$R(T) = R_0 + A_p \cdot \left(\frac{T}{T_\omega}\right)^p \cdot \int_0^{\frac{T_\omega}{T}} \frac{x^p}{(e^x-1)\cdot(1-e^{-x})} \cdot dx \qquad (2)$$

where $T_\omega$ is the characteristic temperature, and $p$ is a free-fitting parameter. It should be noted that for some $R(T)$ curves analysed below we used fixed $p = 5$ value in Eq. 2, and for these cases the designation of $T_\theta$ is kept for the Debye temperature designation.

A primary idea to utilize Eq. 2 to reveal the type of charge carrier interaction is based on well-established theoretical result that $p$ in Eq. 2 approaches unique integer values for different interaction mechanisms [44-46] (see Table I).

**Table I.** *p*-value in generalized Bloch-Grüneisen equation (Eq. 2) and designated for that value an interaction mechanism [44-46].

| *p* | **Interaction mechanism** |
|---|---|
| 2 | the electron-electron |
| 3 | the electron-magnon |
| 5 | the electron-phonon |

Thus, the dominant charge carrier interaction mechanism in given materials can be determined from the comparison of deduced free-fitting parameter *p* with theoretical values for pure cases (Table I).

Because all considered $R(T)$ datasets were measured for superconductors, we used the recently proposed equation [41,43] to fit full $R(T)$ curve, including the superconducting transition:

$$R(T) = R_0 + \theta(T_c^{onset} - T) \cdot \left( \frac{R_{norm}}{\left( I_0\left( F \cdot \left(1 - \frac{T}{T_c^{onset}}\right)^{3/2} \right) \right)^2} \right) + \theta(T - T_c^{onset}) \cdot \Bigg( R_{norm} + A \cdot$$

$$\left( \left(\frac{T}{T_\omega}\right)^p \cdot \int_0^{\frac{T_\omega}{T}} \frac{x^p}{(e^x-1)\cdot(1-e^{-x})} \cdot dx - \left(\frac{T_c^{onset}}{T_\omega}\right)^p \cdot \int_0^{\frac{T_\omega}{T_c^{onset}}} \frac{x^p}{(e^x-1)\cdot(1-e^{-x})} \cdot dx \right) \Bigg) \quad (3)$$

where $T_c^{onset}$ is free-fitting parameter of the onset of superconducting transition, $R_{norm}$ is the sample resistance at the onset of the transition, $\theta(x)$ is the Heaviside step function, $I_0(x)$ is the zero-order modified Bessel function of the first kind and $F$ is a free-fitting dimensionless parameter.

## 3. Results and Discussion

### 3.1. Highly-compressed lithium

Shimizu *et al* [31] (in their Fig. 2) reported $R(T)$ curves for lithium compressed at $P = 3.5$, 23, 35 and 36 GPa. Due to the overlapping of $R(T)$ curves at $P = 35$ and $P = 36$ GPa, in Fig. 1 we fitted $R(T)$ datasets measured at $P = 23$ and $35$ GPa. It can be seen in Fig. 1 that deduced $p$-value for both pressures are remarkably close to each other (i.e. $p = 2.7 - 2.8$).

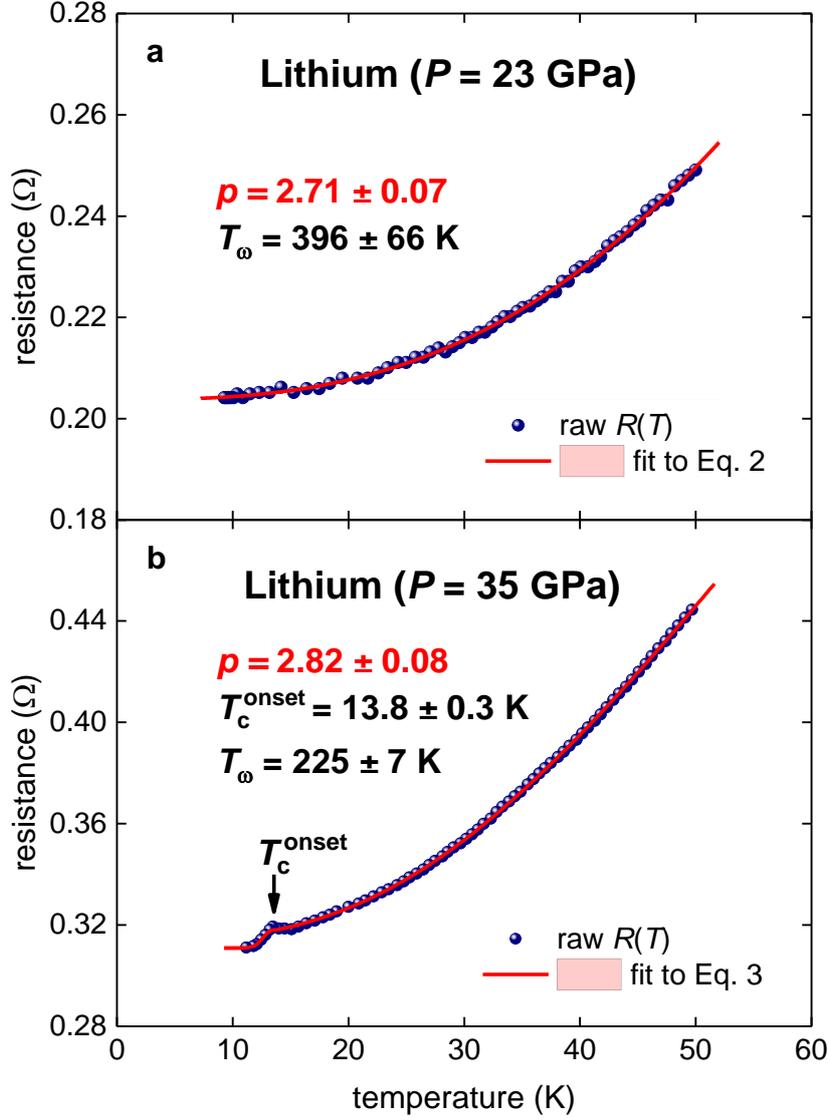

**Figure 1.** Resistance data $R(T)$ and fit to Eq. 2 (a) and Eq. 3 (b) for highly-compressed lithium (raw $R(T)$ data reported by Shimizu *et al* [31]). **a** – deduced $p = 2.71 \pm 0.07$ and $T_\omega = 396 \pm 66\ K$; the fit quality is 0.9990. **b** – deduced $p = 2.82 \pm 0.08$, $T_c^{onset} = 13.8 \pm 0.3\ K$ $T_\omega = 225 \pm 7\ K$; the fit quality is 0.9998. 95% confidence bands are shown by pink shaded areas.

Main result of the analysis, i.e. $p = 2.7 - 2.8$, implies that many-fold disagreement between observed $T_c$ and calculated $T_c$ (by the first-principles calculations [29-37]) has natural explanation that the charge carrier pairing in highly-compressed lithium is not belong the electron-phonon interaction.

## 3.2. Highly-compressed black phosphorous

Shirotani *et al* [47] in their Figure 5 reported $\rho(T)$ curve for black phosphorous compressed at $P = 15$ GPa. In Figure 2 we fitted this dataset to Eq. 3, where in panel (a) *p* was fixed to 5, and in panel (b) *p* was a free-fitting parameter.

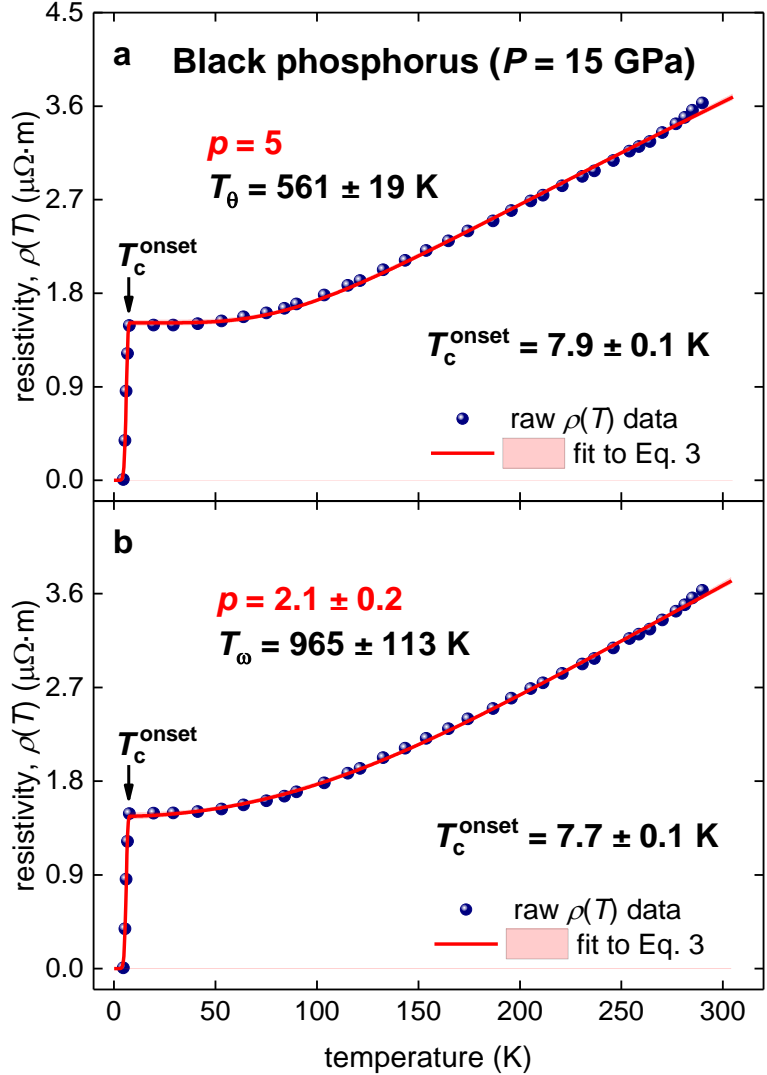

**Figure 4.** Resistivity data, $\rho(T)$, and data fits to Eq. 3 for highly-compressed black phosphorus (raw data is from Ref. 47). (a) $p = 5$, deduced $T_\theta = 561 \pm 19\ K$, fit quality is 0.9986; (b) deduced $p = 2.1 \pm 0.2$, $T_\omega = 965 \pm 113\ K$, fit quality is 0.9992. 95% confidence bands are shown by pink shadow areas.

Despite a fact that the fit at *p* = 5 converged and has high quality and deduced $T_\theta = 561 \pm 19\ K$ implies the weak-coupling scenario in this superconductor (because of $\frac{T_c^{onset}}{T_\theta} \cong 0.01$) even within the electron-phonon pairing mechanism. When *p* is a free-fitting parameter, its deduced value, $p = 2.1 \pm 0.2$, unavoidably means the dominance of the electron-electron interaction in this highly-compressed superconductor.

### 3.3. Highly-compressed sulphur

Yakovlev *et al* [48] reported on the observation of superconductivity in highly-compressed sulphur, which became the first non-conductive element converted into superconductor by applying high pressure. Here in Fig. 3 we fitted temperature dependent resistance data, $\frac{R(T)}{R(T=77\,K)}$, measured at *P* = 76, 86, and 93 GPa by Shimizu *et al* [49] (raw data is shown in Fig. 10 of Ref. 49). These datasets were recently fitted to Eq. 3 at fixed *p* = 5 by Talantsev and Stolze [50] and, thus, fits quality and deduced $T_\theta$ at *p* = 5 can be found in Ref. 50.

Free-fitting parameters *p* and $T_\omega$ deduced from the fits (Fig. 3) are varying in narrow ranges of $p = 2.5 - 2.8$ and of $T_\omega = 319 - 376\,K$. Deduced *p* confidently implies that the charge carrier in highly-compressed sulphur exhibited non-electron-phonon interaction.

### 3.4. Highly-compressed silane

As we already discussed above, highly-compressed silane, SiH$_4$, represents perhaps one of the most challenging case for widely accepted paradigm that the superconductivity in highly-compressed hydrogen-rich compounds originates from the electron-phonon pairing mechanism. In Fig. 4 we showed *R*(*T*) data reported by Eremets *et al* [22] for SiH$_4$ compressed at *P* = 192 GPa (in their Fig. 2(b)) and data fits to Eq. 3 at *p* = 5 (panel a) and *p* being free-fitting parameter (panel b). It can be seen that despite a fact that at *p* = 5 the fit converged and has high quality, when *p* is free-fitting parameter its value is $p = 2.7 \pm 0.2$ and deduced free-fitting characteristic temperature is $T_\omega = 435 \pm 17\,K$.

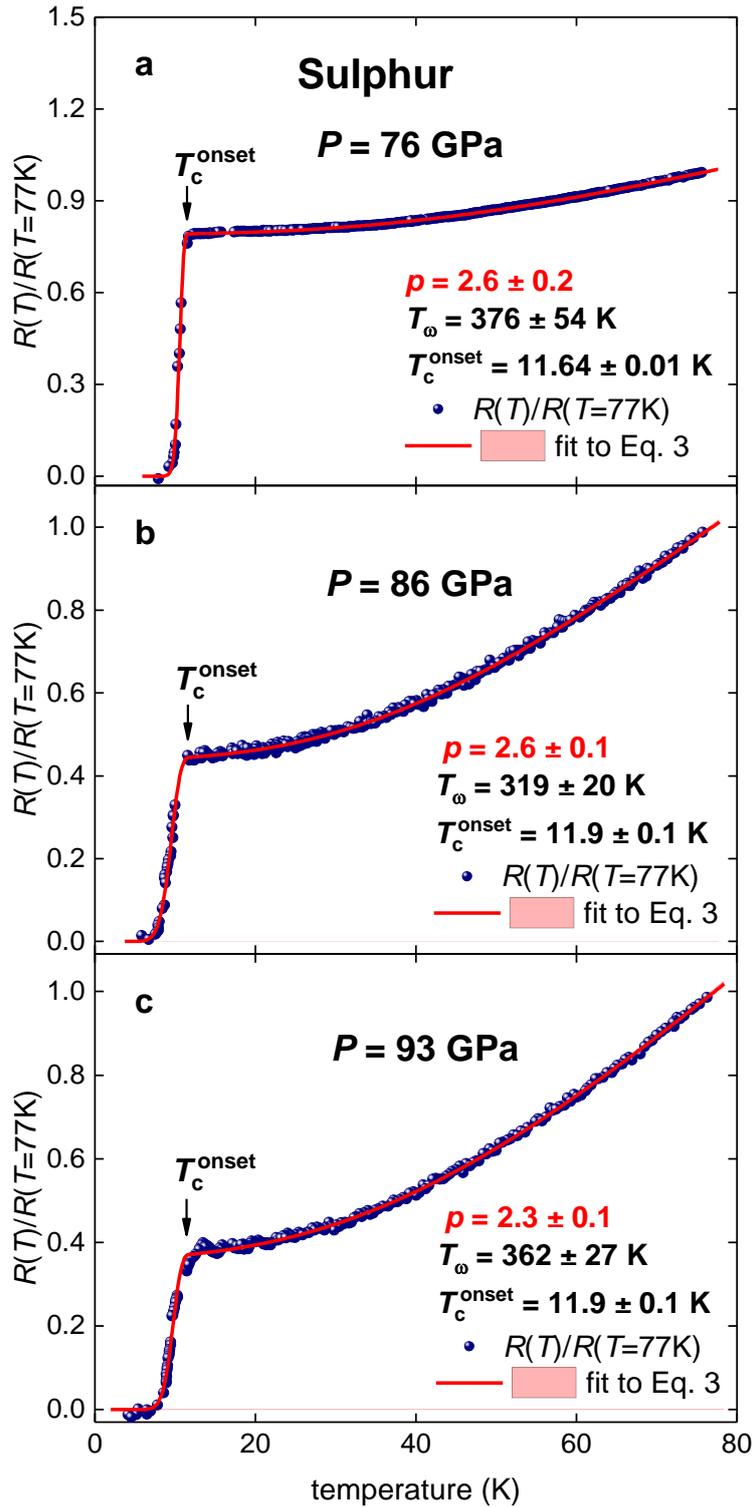

**Figure 3.** $R(T)/R(T=77K)$ datasets for highly-compressed sulfur (raw data is from Ref. 49) and fits to Eq. 3 at $p$ being a free-fitting parameter. (a) $P = 76$ GPa, deduced $p = 2.6 \pm 0.2$, $T_\omega = 376 \pm 54\ K$, fit quality is 0.9979; (b) $P = 86$ GPa, deduced $p = 2.6 \pm 0.1$, $T_\omega = 319 \pm 20\ K$, fit quality is 0.9982; (c) $P = 93$ GPa, deduced $p = 2.3 \pm 0.1$, $T_\omega = 362 \pm 27\ K$, fit quality is 0.9985. 95% confidence bands are shown by pink shadow areas.

The first outcome of our analysis is that neither deduced $T_\theta = 435 \pm 17\ K$, nor $T_\omega = 435 \pm 17\ K$ is closed to calculated by Feng *et al* [21] the Debye temperature of $T_\theta = 3,500 - 4,000\ K$. The second outcome, that free-fitting parameter $p = 2.7 \pm 0.2$ implies the non-electron-phonon charge carrier interaction in this highly-compressed hydrogen-rich compound.

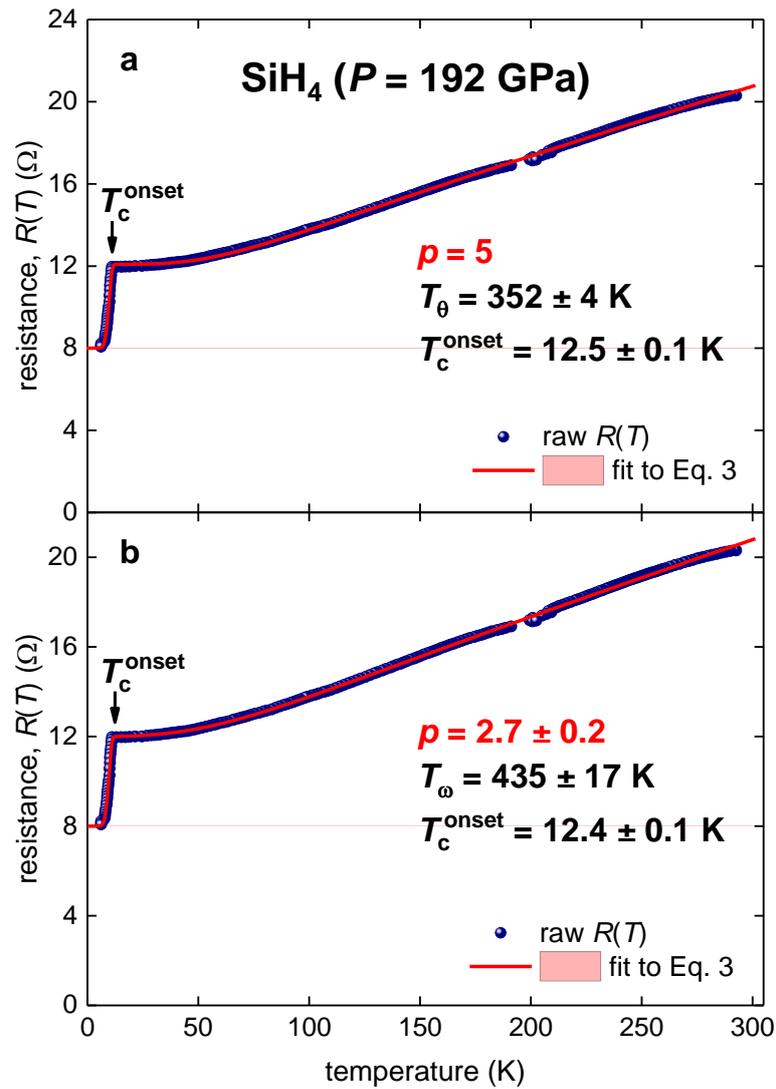

**Figure 4.** $R(T)$ data and fits to Eq. 3 for highly-compressed silane ($P = 192$ GPa) (experimental data digitized from Fig. 2(b) in Ref. 22). (a) $p = 5$, deduced $T_\theta = 352 \pm 4\ K$, fit quality is 0.9995; (b) deduced $p = 2.7 \pm 0.2$, $T_\omega = 435 \pm 17\ K$, fit quality is 0.9996. 95% confidence bands are shown by pink shadow areas.

## 4. Conclusion

In conclusion, in this paper we analysed $R(T)$ data for highly-compressed elemental superconductors lithium, black phosphorous, and sulfur, and also for simple hydrogen-rich silane, $SiH_4$. Overall, all studied superconductors exhibit very close values for parameter $p$ in the generalized Bloch-Grüneisen (BG) equation (Eqs. 2,3) which varies within a narrow range of $p = 2.0 − 2.8$. This range of $p$ is very different from $p = 5$ (i.e., the unique characteristic value belonging the electron-phonon charge carrier interaction). This result is in a good accord with recent report by Talantsev [41], who reported essentially the same deduced values for $p = 1.8 − 3.2$ for highly-compressed boron, $H_3S$, $LaH_x$, $PrH_9$ and $BaH_{12}$.

Overall, this implies that non-electron-phonon mechanisms for the emergence of the superconductivity in highly-compressed materials [40,51-53] should be thorough considered.

## References

[1] Satterthwaite C B and Toepke I L 1970 Superconductivity of hydrides and deuterides of thorium *Phys. Rev. Lett.* **25** 741-743
[2] Ashcroft N W 1968 Metallic hydrogen: a high-temperature superconductor? *Phys. Rev. Lett.* **21** 1748-1749
[3] Drozdov A P, Eremets M I, Troyan I A, Ksenofontov V, Shylin S I 2015 Conventional superconductivity at 203 kelvin at high pressures in the sulfur hydride system *Nature* **525** 73-76
[4] Drozdov A P, *et al* 2019 Superconductivity at 250 K in lanthanum hydride under high pressures *Nature* **569** 528-531
[5] Somayazulu M, *et al* 2019 Evidence for superconductivity above 260 K in lanthanum superhydride at megabar pressures *Phys. Rev. Lett.* **122** 027001
[6] Zhou D, *et al* 2020 Superconducting praseodymium superhydrides *Science Advances* **6** eaax6849
[7] Chen W, *et al* 2021 Synthesis of molecular metallic barium superhydride: pseudocubic $BaH_{12}$ *Nature Communications* **12** 273
[8] Wang N, *et al* 2021 A low-$T_c$ superconducting modification of $Th_4H_{15}$ synthesized under high pressure *Superconductor Science and Technology* **34** 034006
[9] Troyan I A, *et al* 2021 Anomalous high-temperature superconductivity in $YH_6$ *Adv. Mater.* **33** 2006832
[10] Kong P P, *et al* 2019 Superconductivity up to 243 K in yttrium hydrides under high pressure (arXiv:1909.10482)
[11] Ma L, *et al* 2021 Experimental observation of superconductivity at 215 K in calcium superhydride under high pressure (arXiv:2103.16282)

[12] Li Z W, *et al* 2021 Superconductivity above 200 K observed in superhydrides of calcium (arXiv:2103.16917)
[13] Hong F, *et al* 2021 Superconductivity at ∼70 K in tin hydride SnHx under high pressure (arXiv:2101.02846)
[14] Semenok D V, *et al* 2020 Superconductivity at 253 K in lanthanum–yttrium ternary hydrides (arXiv:2012.04787)
[15] Zhou D, *et al* 2020 Superconducting praseodymium superhydrides *Sci. Adv.* **6** eaax6849
[16] Chen W, *et al* 2021 High-pressure synthesis of barium superhydrides: Pseudocubic $BaH_{12}$ *Nat. Commun.* **12** 273
[17] Drozdov A P, Eremets M I and Troyan I A 2015 Superconductivity above 100 K in $PH_3$ at high pressures (arXiv:1508.06224)
[18] Matsuoka T, *et al* 2019 Superconductivity of platinum hydride *Phys. Rev. B* **99** 144511
[19] Chen W, Semenok D V, Huang X, Shu H, Li X, Duan D, Cui T and Oganov A R 2021 High-temperature superconductivity in cerium superhydrides (arXiv:2101.01315)
[20] Semenok D V et al 2020 Superconductivity at 161 K in thorium hydride $ThH_{10}$: synthesis and properties *Mater. Today* **33** 36–44
[21] Feng J, *et al* 2006 Structures and potential superconductivity in $SiH_4$ at high pressure: En route to "metallic hydrogen" *Phys. Rev. Lett.* **96** 017006
[22] Eremets M I, Trojan I A, Medvedev S A, Tse J S, Yao Y 2008 Superconductivity in hydrogen dominant materials: Silane *Science* **319** 1506-1509
[23] Talantsev E F 2020 Advanced McMillan's equation and its application for the analysis of highly-compressed superconductors *Superconductor Science and Technology* **33** 094009
[24] Bloch F 1930 Zum elektrischen Widerstandsgesetz bei tiefen Temperaturen *Z. Phys.* **59** 208-214
[25] Grüneisen E 1933 Die abhängigkeit des elektrischen widerstandes reiner metalle von der temperatur. *Ann. Phys.* **408** 530–540
[26] Mozaffari S, *et al* 2019 Superconducting phase diagram of $H_3S$ under high magnetic fields *Nature Communications* **10** 2522
[27] Yao Y, Tse J S, Ma Y and Tanaka K 2007 Superconductivity in high-pressure $SiH_4$ *EPL* **78** 37003
[28] Chen X-J, *et al* 2008 Pressure-induced metallization of silane *PNAS* **105** 20-23
[29] Christensen N E, Novikov D L 2001 Predicted superconductive properties of lithium under pressure *Phys. Rev. Lett.* **86** 1861-1864
[30] Lin T H and Dunn K J 1986 High-pressure and low-temperature study of electrical resistance of lithium *Phys. Rev. B* **33** 807-811
[31] Shimizu K, Ishikawa H, Takao D, Yagi T and Amaya K 2002 Superconductivity in compressed lithium at 20 K *Nature* **419** 597-599
[32] Struzhkin V V, Eremets M I, Gan W, Mao H-K, Hemley R J 2002 Superconductivity in dense lithium *Science* **298** 1213-1215
[33] Shimizu K, Amaya K and Suzuki N 2005 Pressure-induced superconductivity in elemental materials *Journal of the Physical Society of Japan* **74** 1345-1357
[34] Deemyad S and Schilling J S 2003 Superconducting phase diagram of Li metal in nearly hydrostatic pressures up to 67 GPa *Phys. Rev. Lett.* **91** 167001
[35] Matsuoka T and Shimizu K 2009 Direct observation of a pressure-induced metal-to-semiconductor transition in lithium
[36] Schaeffer A M, Temple S R, Bishop J K, and Deemyad S 2015 High-pressure superconducting phase diagram of $^6Li$: Isotope effects in dense lithium *PNAS* **112** 60-64
[37] Profeta G, *et al* 2006 Superconductivity in lithium, potassium, and aluminum under extreme pressure: A first-principles study *Phys. Rev. Lett.* **96** 047003

[38] Bardeen J, Cooper L N, and Schrieffer J R 1957 Theory of superconductivity *Phys. Rev.* **108** 1175-1204
[39] Eliashberg G M 1960 Interactions between electrons and lattice vibrations in a superconductor *Soviet Phys. JETP* **11** 696-702
[40] Kim H-T 2021 Room-temperature-superconducting $T_c$ driven by electron correlation *Scientific Reports* **11** 10329
[41] Talantsev E F 2021 The dominance of non-electron-phonon charge carrier interaction in highly-compressed superhydrides *arXiv*:2104.14145
[42] Jiang H, *et al* 2015 Physical properties and electronic structure of $Sr_2Cr_3As_2O_2$ containing $CrO_2$ and $Cr_2As_2$ square-planar lattices *Phys. Rev. B* **92** 205107
[43] Talantsev E 2021 Quantifying the charge carrier interaction in metallic twisted graphene superlattices *Nanomaterials* **11** 1306
[44] White G K and Woods S B 1959 Electrical and thermal resistivity of the transition elements at low temperatures *Phil. Trans. R. Soc. Lond. A* **251**, 273-302
[45] Matula R A 1979 Electrical resistivity of copper, gold, palladium, and silver *J. Phys. Chem. Ref. Data* **8** 1147-1298
[46] Poker D B and Klabunde C E 1982 Temperature dependence of electrical resistivity of vanadium, platinum, and copper *Phys. Rev. B* **26** 7012
[47] Shirotani I, *et al* 1994 Phase transitions and superconductivity of black phosphorus and phosphorus-arsenic alloys at low temperatures and high pressures *Phys. Rev. B* **50** 16274-16278
[48] Yakovlev E N, Stepanov G N, Timofeev Y A and Vinogradov B V 1978 Superconductivity of sulphur at high pressure *JETP Lett.* **28** 340–342
[49] Shimizu K, Amaya K and Suzuki N 2005 Pressure-induced superconductivity in elemental materials *J. Phys. Soc. Japan* **74** 1345-1357
[50] Talantsev E F and Stolze K 2021 Resistive transition in hydrogen-rich superconductors *Superconductor Science and Technology* **34** 064001
[51] Monthoux P, Pines D and Lonzarich G G 2007 Superconductivity without phonons *Nature* **450** 1177-1183
[52] Sachdev S 2012 What can gauge-gravity duality teach us about condensed matter physics? *Annual Review of Condensed Matter Physics* **3** 9-33
[53] Matthias B T 1971 Anticorrelations in superconductivity *Physica* **55** 69-72